\documentstyle[aps,prd,preprint,epsfig,tighten]{revtex}
\begin{document}
\draft
\title{Model-dependent and independent implications \\
of the first Sudbury Neutrino Observatory results}
\author{	G.L.\ Fogli$^a$, 
		E.\ Lisi$^a$, 
		D.\ Montanino$^b$, 
		and A.\ Palazzo$^a$\\[4mm]
}
\address{$^a$Dipartimento di Fisica and Sezione INFN di Bari\\
             	Via Amendola 173, 70126 Bari, Italy \\ }
\address{$^b$Dipartimento di Scienza dei Materiali and Sezione INFN di Lecce\\
             	Via Arnesano, 73100 Lecce, Italy \\ }
\maketitle
\begin{abstract}
We briefly discuss some implications of the first solar $\nu$ results from the
Sudbury Neutrino Observatory (SNO) experiment in the charged-current channel.  
We first show that the present SNO response  function is very similar to the
Super-Kamiokande (SK) one above 8.6 MeV  in kinetic electron energy. On the
basis of such equivalence we confirm, in a completely  model-independent way,
the SNO evidence for an active,  non-electron neutrino component in the SK
event sample, with a significance greater than $3\sigma$.  Then, by assuming no
oscillations into sterile neutrinos, we combine the SK+SNO data  to derive
allowed regions for two free parameters: (i) the ratio $f_B$ of the {\em true}
$^8$B $\nu$ flux from the Sun to the corresponding value predicted by the
standard solar model (SSM),  and (ii) the $\nu_e$ survival probability 
$\langle P_{ee}\rangle$, averaged over the common SK and SNO response function.
We obtain the separate $3\sigma$ ranges: $f_B=1.03^{+0.50}_{-0.58}$ (in
agreement with the SSM central value, $f_B=1$) and $\langle P_{ee}\rangle
=0.34^{+0.61}_{-0.18}$ (in $>3\sigma$ disagreement with the standard
electroweak model prediction, $\langle P_{ee}\rangle=1$), with strong
anticorrelation between the two parameters. Finally, by taking $f_B$ and its
uncertainties as predicted by the SSM, we perform an updated analysis of the
$2\nu$ active neutrino oscillation parameters $(\delta m^2,\tan^2\omega)$
including all the solar $\nu$ data, as well as the spectral data from the CHOOZ
reactor experiment. We find that only the solutions at $\tan^2\omega\sim
O(1)$   survive at the $3\sigma$ level in the global fit,  with a  preference
for the one at high  $\delta m^2$---the so-called large mixing angle solution. 
\end{abstract}
\medskip

\section{Introduction}

The Sudbury Neutrino Observatory (SNO) experiment \cite{SNOe}  has recently
presented the first measurements of the $\nu_e+d\to p+n+e^-$ reaction rate
induced by $^8$B solar neutrinos through charged currents (CC) \cite{SNOc} . 
The observed CC event rate, normalized to the latest standard solar model (SSM)
prediction \cite{BP00},
\begin{equation}
{\rm SNO/SSM}=0.347 \pm 0.029\ ,
\label{SNO}
\end{equation}
not only confirms the deficit of solar neutrino events previously observed by
the chlorine \cite{Cl00}, gallium \cite{Ga00},  and water-Cherenkov
\cite{Kami,SK01}  experiments, but provides a $>3\sigma$ evidence \cite{SNOc}
for a $\nu_{\mu,\tau}$  contribution in the Super-Kamiokande (SK) measurement
of the  $\nu_x+e^-\to \nu_x+e^-$ reaction rate ($x=e,\mu,\tau$)  in a similar
energy range \cite{SK01},
\begin{equation}
{\rm SK/SSM}=0.459 \pm 0.017\ .
\label{SK}
\end{equation}

The SK-SNO comparison can be made rigorously model-independent by making an
appropriate choice for the SK energy threshold, as suggested in
\cite{Res1,Res2} and discussed also in the SNO paper  \cite{SNOc}. In this work
we first provide (Sec.~II)  an improved discussion of such model-independent
comparison, which, based on the (previously undisclosed) detailed SNO detector
specifications, confirms the $>3\sigma$ evidence for a $\nu_{\mu,\tau}$  flavor
component in the SK event sample. We then assume (Sec.~III)  no oscillations
into sterile neutrinos, and derive combined constraints on  two free
parameters: (i) the ratio $f_B$ of the {\em true} $^8$B $\nu$ flux from the Sun
to the corresponding value predicted by the SSM, and (ii) the $\nu_e$ survival
probability  $\langle P_{ee}\rangle$ averaged over the SK-SNO response
function. Such constraints confirm the SSM prediction for $f_B$, and strongly
indicate an average $\nu_e$ flux suppression of about one third  $(\langle
P_{ee}\rangle\sim 1/3$).  Finally (Sec.~IV) we assume the validity of the SSM,
and perform and updated analysis of all the available solar neutrino data
(including the SNO event rate) in a $2\nu$ active oscillation framework. Large
mixing angle solutions are clearly preferred in the global fit, while the
small-mixing one is not allowed at the $3\sigma$ level (99.73\% C.L.)  by the
parameter estimation test. We conclude our work in Sec.~V.

\section{Using the SK-SNO equivalence (with no additional assumption)}

An important characteristic of any solar neutrino experiment is the energy
spectrum of parent neutrinos contributing to the collected event sample in the
absence of oscillations---the so-called response function $\varrho(E_\nu)$ 
\cite{Faid}. The response function basically folds the solar neutrino energy
spectrum with both the differential  interaction cross section and with the
detector threshold and energy resolution, and thus it takes different  forms
for each experiment. However, it can be made (partly accidentally) equal in SK
and SNO by an appropriate choice of the detected electron energy thresholds
(or, more generally, energy ranges), as shown in \cite{Res1,Res2}  on the basis
of the {\em expected\/} SNO technical specifications. 

By repeating the analysis in \cite{Res1}  with the present SNO kinetic energy
threshold $(T_e^{\rm SNO}\geq 6.75$ MeV) and energy resolution \cite{SNOc}, we
find a best-fit SK-SNO agreement for a SK threshold  $T_e^{\rm SK}\geq 8.6$ MeV
(instead of $T^{\rm SK}_{e}>5{\rm\ MeV}-m_e$, for which the value in 
Eq.~(\ref{SK}) is officially quoted \cite{SK01}). Such ``adjusted'' threshold
is in good agreement with the one estimated in the SNO paper ($T_{e}^{\rm
SK}\geq 8.5$ MeV \cite{SNOc}).  Figure~1 displays our calculations for the
corresponding SK and SNO response functions to $^8$B neutrinos, which appear to
be in very good agreement with each other. Concerning  the small shape
difference in the first half of the response functions, we estimate that, in
our analysis, the corresponding effect is (in the worst case) a factor of five
smaller than the effect of the total SNO uncertainty in Eq.~(\ref{SNO}), so
that we can safely take $\varrho_{\rm SK}=\varrho_{\rm SNO}$. With the adjusted
SK threshold, the SK and SNO detectors are thus equally sensitive to the
incoming $^8$B neutrinos. The possible small contribution of {\em hep\/}
neutrinos (not shown in Fig.~1) does not spoil the SK-SNO equalization of
response functions  \cite{Res2}, as far as the {\em hep\/} flux is taken below
the  experimental upper limit provide by the latest SK spectral measurements
\cite{SK01}.

Although there is no official number quoted yet by the SK collaboration for the
SK/SSM value at $T_{e}^{\rm SK}\geq 8.6$ MeV, one can try to recover it from
the published SK spectral data  \cite{SK01,SKsp}. We adopt the provisional SNO
own estimate \cite{SNOc}, corresponding to take 
\begin{equation}
{\rm SK/SSM}=0.451 \pm 0.017 \;\; (T_{e}^{\rm SK}\geq 8.6{\rm\ MeV})\ ,
\label{SK'}
\end{equation}
which amounts to a small shift in the central value of the total SK rate.  The
attached SK error is assumed to be basically the same as in Eq.~(\ref{SK}),
since it should be dominated by systematic errors rather than by statistical
uncertainties. Furthermore, the SK-SNO comparison is dominated by the
(presently) larger SNO uncertainties, so that any (presumably small) official
SK re-evaluation of the  numbers in Eq.~(\ref{SK'}) is not expected to produce
significant changes  in the results discussed below \cite{MSmy}.

For $\varrho_{\rm SK}=\varrho_{\rm SNO}$, the following relations hold {\em
exactly\/} \cite{Res1,Res2}:
\begin{eqnarray}
{\rm SNO/SSM} &=& f_B \langle P_{ee}\rangle\ ,
\label{SNOrel}\\
{\rm SK/SSM} &=& f_B \langle P_{ee}\rangle + f_B \frac{\sigma_a}{\sigma_e}
\langle P_{ea} \rangle\ ,
\label{SKrel}
\end{eqnarray}
where $f_B$ is the ratio between the true (unknown) $^8$B $\nu_e$ flux at the
Sun and its SSM prediction \cite{BP00},  $\langle P_{ee}\rangle$ is the $\nu_e$
survival probability (energy-averaged over the common SK-SNO response
function), $\langle P_{ea}\rangle$ is the averaged transition probability to
active neutrinos $(\nu_a=\nu_{\mu,\tau})$, and $\sigma_a/\sigma_e$ is the ratio
of the  (properly averaged \cite{Res1,Res2}) cross sections of $\nu_a$ and
$\nu_e$ on electrons.  We calculate $\sigma_a/\sigma_e=0.152$  for $T_e^{\rm
SK}\geq 8.6$~MeV. Notice that the above relations do not imply any assumption
either on $f_B$, or  on possible sterile neutrino oscillations, or on the
functional form of $P_{ee}(E_\nu)$ or $P_{ea}(E_\nu)$, and thus they are {\em
completely model-independent}.

From the above relations one can derive that:
\begin{eqnarray}
{\rm SK/SSM} &<& {\rm SNO/SSM}\;\;\; {\rm is\ always\ forbidden\ } 
				(\langle P_{\mu\tau}\rangle <0)\ ,\\
{\rm SK/SSM} &=& {\rm SNO/SSM}\;\;\; {\rm is\ allowed\ only\ if\ } 
				\langle P_{\mu\tau}\rangle =0\ ,\\
{\rm SK/SSM} &>& {\rm SNO/SSM}\;\;\; {\rm is\ allowed\ only\ if\ } 
				\langle P_{\mu\tau}\rangle >0\ .
\end{eqnarray}
Figure~2 displays the above constraints at a glance. The SK+SNO experimental
data are well within  the region where there   {\em must\/} be
$\nu_e\to\nu_{\mu,\tau}$ transitions, independently of a possibly open
\cite{Barg} $\nu_e\to\nu_s$ channel. Only at $>3\sigma$ (more precisely, at
$3.1$ sigma) the experimental data would hit the diagonal line which
parametrizes the case of no $\nu_e\to\nu_{\mu,\tau}$ transitions (corresponding
to either no oscillations or pure $\nu_e\to\nu_s$ oscillations). Such results
represent an alternative way to look at the SNO evidence \cite{SNOc} for a
$\nu_{\mu,\tau}$ component in the SK events at $>3\sigma$.

\section{Using the SK-SNO equivalence (without sterile neutrinos)}

The SNO results \cite{SNOe} and the model-independent analysis in the previous
section show that there is evidence for active neutrino transitions. Therefore,
it is legitimate to explore the consequences of the {\em additional\/}
hypothesis of {\em purely active\/} flavor transitions, corresponding to take
$\langle P_{ea}\rangle =1-\langle P_{ee}\rangle$.  In such a case, the SK-SNO
relations in Eqs.~(\ref{SNOrel}) and (\ref{SKrel}) read
\begin{eqnarray}
{\rm SNO/SSM} &=& f_B \langle P_{ee}\rangle\ ,
\label{SNOrel2}\\
{\rm SK/SSM} &=& f_B \langle P_{ee}\rangle + f_B \frac{\sigma_a}{\sigma_e}
(1-\langle P_{ee} \rangle)\ ,
\label{SKrel2}
\end{eqnarray}
providing a system of two equations in the two unknowns $f_B$ and  $\langle
P_{ee}\rangle$. By fitting the experimental values of SNO/SSM and SK/SSM given
in Eqs~(\ref{SNO}) and (\ref{SK'}), respectively, one can then determined
allowed ranges for $f_B$ and  $\langle P_{ee}\rangle$.

Figure~3 shows the contours of the allowed region in the ($\langle
P_{ee}\rangle,\,f_B$)  plane for $\chi^2=\Delta\chi^2=1,4,$ and 9, whose {\em
projections} onto the coordinate axes give the $1\sigma$, $2\sigma$, and
$3\sigma$ separate ranges \cite{PDBS} for $f_B$ and $\langle P_{ee}\rangle$.
The strong anticorrelation  reflects the fact that a high $^8$B flux can be
partly compensated by a smaller survival probability, and vice versa.   The
projected $3\sigma$ range for $f_B$ ($f_B=1.03^{+0.50}_{-0.58}$) is in
agreement with the SSM  prediction   (shown with its $\pm1\sigma$ error band
from \cite{BP00},  $f_B=1^{+0.20}_{-0.16}$).  On the other hand,  the projected
$3\sigma$ range for the average survival probability ($\langle P_{ee}\rangle
=0.34^{+0.61}_{-0.18}$), clashes  with the standard electroweak model
prediction of electron flavor conservation ($\langle P_{ee}\rangle=1$) at
$>3\sigma$. In the context of this figure, the standard model of the Sun 
appears to be in better shape that the standard model of electroweak
interactions.

The results indicate that, in the case of generic active oscillations (i.e., no
other assumption apart from $\langle P_{es}\rangle=0$ in the range probed
jointly by SK and SNO), the $\nu_e$ survival probability takes basically the
lowest value allowed by pre-SNO experiments. It has been shown in \cite{BaKr}
that the lowest values of  $\langle P_{ee}\rangle$ (in the $^8$B energy range)
are typically reached within the so-called large mixing angle (LMA) solution to
the solar neutrino problem, which may therefore be expected as favored. This
will be confirmed by the analysis in the next section.

\section{Using all the solar neutrino data (assuming the SSM and
two-family active
oscillations)}

From the SNO results \cite{SNOc} and from the analysis in the previous Sections
we have learned that: (i) There is evidence for active neutrino oscillations,
and (ii) Assuming purely active $\nu$ oscillations, the SSM is confirmed and
the $\nu_e$ survival probability should be $\sim 1/3$ in the SK-SNO energy
range. Let us now make two further assumptions about neutrino physics, namely,
that the $\nu$ fluxes from the Sun can be taken as predicted (with their
uncertainties) by the SSM \cite{BP00}, and that active neutrino oscillations
occur in an effective two-family framework. The latter hypothesis  is totally
correct if the mixing angle $\theta_{13}$ vanishes, and  is accurate up to
$O(\sin^2\theta_{13})$ corrections if $\theta_{13}>0$.  We remind that the
joint analyses of SK atmospheric neutrino data \cite{SKat} and of the CHOOZ
reactor results \cite{CHOO} place stringent upper limits on $\theta_{13}$
\cite{3atm,34at,Vall}.  Moreover, the CHOOZ data forbid large $\nu_e$
disappearance for neutrino square mass differences higher than  $\sim 0.7
\times 10^{-3}$ eV$^2$, and thus they are also relevant to cut away the region
of  energy-averaged solar neutrino oscillations \cite{Vall,Mori,Qave}. 
Therefore, we perform a $2\nu$ analysis by adding the final CHOOZ spectral
results \cite{CHOO} (14 bin, as discussed in \cite{Qave})  to the usual solar
neutrino data, and show then the results in the mass-mixing plane $(\delta
m^2,\tan^2\omega)$, covering both octants in $\omega=\theta_{12}$ \cite{Octa}. 
Since we are now assuming a specific functional form for $P_{ee}(E_\nu)$ (i.e.,
the one predicted by standard oscillation theory at any given mass-mixing
point), the SK-SNO model-independent comparison becomes unimportant, and we can
use the full SK rate given in Eq.~(\ref{SK}) \cite{SK01}.

Concerning SNO, in this work we  include the total CC rate [Eq.~(\ref{SNO})]
but not the published CC energy spectrum  \cite{SNOc}. Notice that the present
SNO spectrum information should be subdominant as compared to SK since,
although one expects a SNO sensitivity to spectral deviations a factor of two
larger than in SK \cite{SKSN}, the current SNO spectral errors (both
statistical and systematic) are more than a factor of two larger than in SK
(and the published SNO event sample is an order of magnitude smaller than the
SK one). Moreover, it is not easy to recover (from both the published SNO and
SK data) the information needed to propagate {\em jointly\/}, on both the SK
and SNO spectra, the correlated $^8$B shape spectrum uncertainties \cite{BaLi}
around the best-fit $\nu$ spectrum (that we take from \cite{Orti} as in
\cite{SK01}). Therefore, we  prefer to postpone the SNO spectrum analysis (and
its properly correlated combination with the SK spectrum) to a future work
\cite{FUTU}. The total SNO CC rate is, however, already important by itself,
and to appreciate its impact  we show first the $2\nu$ analysis {\em without\/}
SNO for reference.

Figure~4 shows the results of the $2\nu$ analysis  using the three pre-SNO
total solar neutrino rates (chlorine \cite{Cl00}, combined gallium
\cite{Ga00},  SK \cite{SK01}) and to the 14-bin CHOOZ data \cite{CHOO}
(relevant to suppress the likelihood of the high-$\delta m^2$ region), as
derived by drawing iso-$\Delta\chi^2$ contours (for $N_{\rm DF}=2$)  around the
global $\chi^2$ minimum. The fit in Fig.~4 favors the small-mixing angle (SMA)
solution, as compared to the regions at $\tan^2\omega\sim O(1)$, usually
referred to as large-mixing angle at high $\delta m^2$ (LMA) and at low $\delta
m^2$ (LOW), extending down to the  quasivacuum and vacuum oscillation (QVO and
VO) regions.   Figure~5 shows the impact of the SK day-night spectral data
\cite{SK01,SKsp} (19+19 bins minus one adjustable normalization factor),  that
cut away the vacuum solutions and also change the relative likelihood of  the
local SMA, LMA, and LOW best fits, favoring the LMA solution (see also
Table~I). Notice also the small region allowed at 99.73\% C.L.\ in the lowest
$\delta m^2$ decade (the so-called Just-So$^2$ solution, see the first of
Ref.~\cite{BaKr} and references therein). Similar results have been largely
discussed  in the recent solar neutrino literature (see, e.g.,
\cite{BaKr,SKsp,Conc,Ours}),  and we do not add further comments here.

Figure~6 is analogous to Fig.~4, but including the SNO CC rate \cite{SNOc}. The
LMA (SMA) solution in Fig.~6 is enlarged (reduced)  as compared to Fig.~4, due
to the anticipated SNO preference for relatively small values of the $\nu_e$
average survival probability, which tend to favor the LMA case. The SMA
solution tends to adapt to the low value  $\langle P_{ee}\rangle\sim 1/3$ by
privileging its rightmost part  (where the nonadiabatic $\nu_e$ suppression is
stronger), and indeed the final compromise makes the SMA local fit comparable
to the LMA one (see Table~I). However, in doing so, the SMA fit also privileges
the part where spectral deviations are sizable, contrary to the SK day-night
spectrum observations.

Indeed, the SMA solution disappears at $>3\sigma$ when the SK day-night
spectral data are included, as shown in Fig.~7 (analogous to Fig.~5, but
including the SNO total CC rate). The ``tension'' between the total rate
information (pushing the SMA to the right) and the SK spectrum (pushing the SMA
to the left), which was already emerging from the latest SK data  analysis
\cite{SKsp}, is now sufficiently strong  to produce a significant  decrease of
the likelihood of the SMA solution. Since the SNO spectral data \cite{SNOc}
(not included here) do not show any deviation from the standard shape within
the (now large) errors,  we may expect that the addition of such data in future
analyses can only corroborate such trend. The LMA solution appears to be
favored in the global fit, enhancing the hopes of interesting new physics at
KamLand \cite{KamL} and at future neutrino factories \cite{Nufa}. The LOW
solution turns out to be slightly less favored then the LMA one,  essentially
because the  gallium data prefer an increase of the $\nu_e$ survival
probability at low energies, which is more easily provided in the LMA region
rather than in the LOW solution (see, e.g., \cite{BaKr}). However, the LOW
solution is still in good shape, and should be tested through day-night earth
matter effects in the BOREXINO experiment \cite{BORE} or, with less
sensitivity, through winter-summer matter effects  after several years of data
taking in the Gallium  Neutrino Observatory (GNO) \cite{GNOB}. Notice that the
LOW solution extends down to the quasivacuum oscillation \cite{Qvac} region,
which might be probed in BOREXINO by pushing its time-variation sensitivity
close to its upper limits \cite{Mura}. Notice that no VO or Just-So$^2$
solutions survive in Fig.~7. Finally, we remark  that the indications for a
relatively small value of  $\langle P_{ee}\rangle$ suggest that a large,
unmistakable neutral-to-current ratio enhancement  (roughly $\propto1/\langle
P_{ee}\rangle$) should be found by the SNO experiment in its second phase of
operation (neutral current mode).

\section{Conclusions}

We have discussed the following implications of the first SNO results (with
increasing degree of model dependence): (i) Evidence for
$\nu_e\to\nu_{\mu,\tau}$ transitions from a  model-independent comparison of SK
and SNO; (ii) Bounds on the $^8$B neutrino flux factor $f_B$  and on the
average $\nu_e$ survival probability under the hypothesis of (generic) active
$\nu$ oscillations; and finally (iii) Marked preference for large-mixing
solutions {\em vs\/} the small mixing one, by assuming both active oscillations
and standard solar model predictions. It seems that the SNO experiment  has
just started to delight us with the first of a series of interesting results.

\begin{table}
\caption{$2\nu$ active oscillations: Positions and local values for the
relevant  $\chi^2$ minima (SMA, LMA, LOW, and Q(VO) solutions). Upper part:
pre-SNO situation, without and with SK day-night spectral data 
\protect\cite{SK01} (19+19 bin).  Lower part: post-SNO situation  (total SNO CC
rate included \protect\cite{SNOc}, 1 datum). In all cases, the fit includes the
chlorine \protect\cite{Cl00},  combined gallium \protect\cite{Ga00} and SK
\protect\cite{SK01} total rates (3 data), as well as the final CHOOZ spectral
data  \protect\cite{CHOO} (14 bin).}
\smallskip
\begin{tabular}{c|ccc|ccc}
& $\log_{10}(\tan^2\omega)$ & $\log_{10}(\delta m^2/{\rm eV}^2)$ & $\chi^2$
& $\log_{10}(\tan^2\omega)$ & $\log_{10}(\delta m^2/{\rm eV}^2)$ & $\chi^2$\\ 
\tableline\tableline
& \multicolumn{3}{c|}{Data: pre-SNO rates + CHOOZ} 
&  \multicolumn{3}{c}{Data: pre-SNO rates + SK spec.\ + CHOOZ} \\
\tableline
SMA   & $-3.03$   & $-5.04$ & 7.70 & $-3.40$ & $-5.10$ & 49.3 \\
LMA   & $-0.54$   & $-4.56$ & 10.6 & $-0.48$ & $-4.31$ & 42.2 \\
LOW   & $-0.14$   & $-7.00$ & 15.6 & $-0.12$ & $-6.99$ & 46.8 \\
(Q)VO & $\pm0.25$ & $-10.0$ & 7.80 & $+0.39$ & $-9.34$ & 48.1 \\
\tableline
& \multicolumn{3}{c|}{Data: post-SNO rates + CHOOZ} 
&  \multicolumn{3}{c}{Data: post-SNO rates + SK spec.\ + CHOOZ} \\
\tableline
SMA   & $-2.94$   & $-5.00$ & 12.0 & $-3.50$ & $-5.10$ & 57.7 \\
LMA   & $-0.35$   & $-4.36$ & 11.7 & $-0.43$ & $-4.31$ & 43.0 \\
LOW   & $-0.20$   & $-6.99$ & 16.4 & $-0.17$ & $-6.97$ & 47.5 \\
(Q)VO & $\pm0.53$ & $-10.1$ & 10.5 & $+0.31$ & $-9.32$ & 49.0 \\
\end{tabular}
\label{tablefluxes}
\end{table}

%
\newcommand{\InsertFigure}[2]{\newpage\begin{center}\mbox{%
\epsfig{bbllx=1.4truecm,bblly=1.3truecm,bburx=19.5truecm,bbury=26.5truecm,%
height=20truecm,figure=#1}}\end{center}\vspace*{-1.8truecm}%
\parbox[t]{\hsize}{\small\baselineskip=0.5truecm\hspace*{0.5truecm} #2}}
\InsertFigure{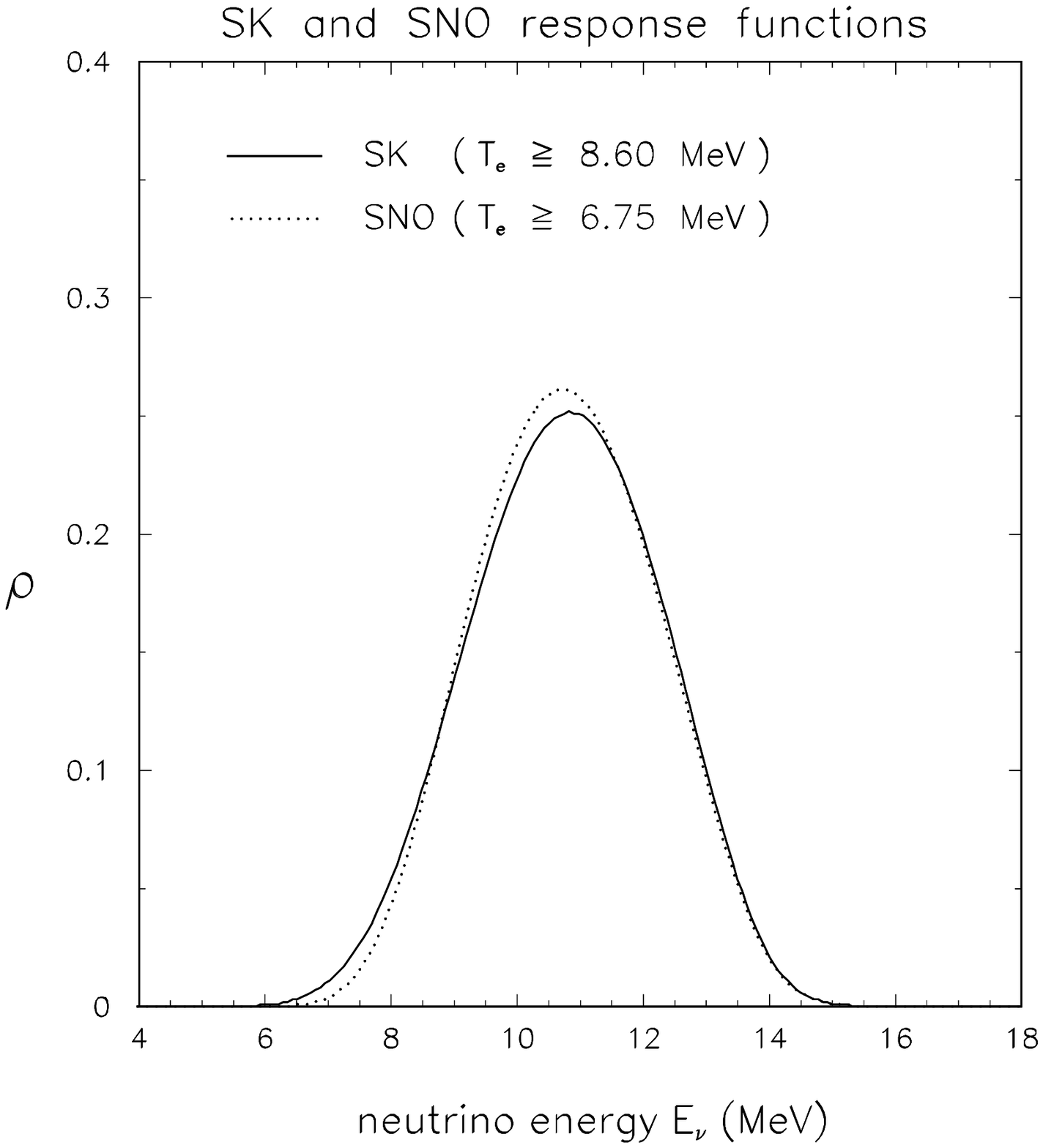}%
{Fig.~1. Best-fit equalization of the SK and SNO response
functions to $^8$B neutrinos, as obtained  by shifting the SK threshold  to 8.6
MeV in electron kinetic energy.}
\InsertFigure{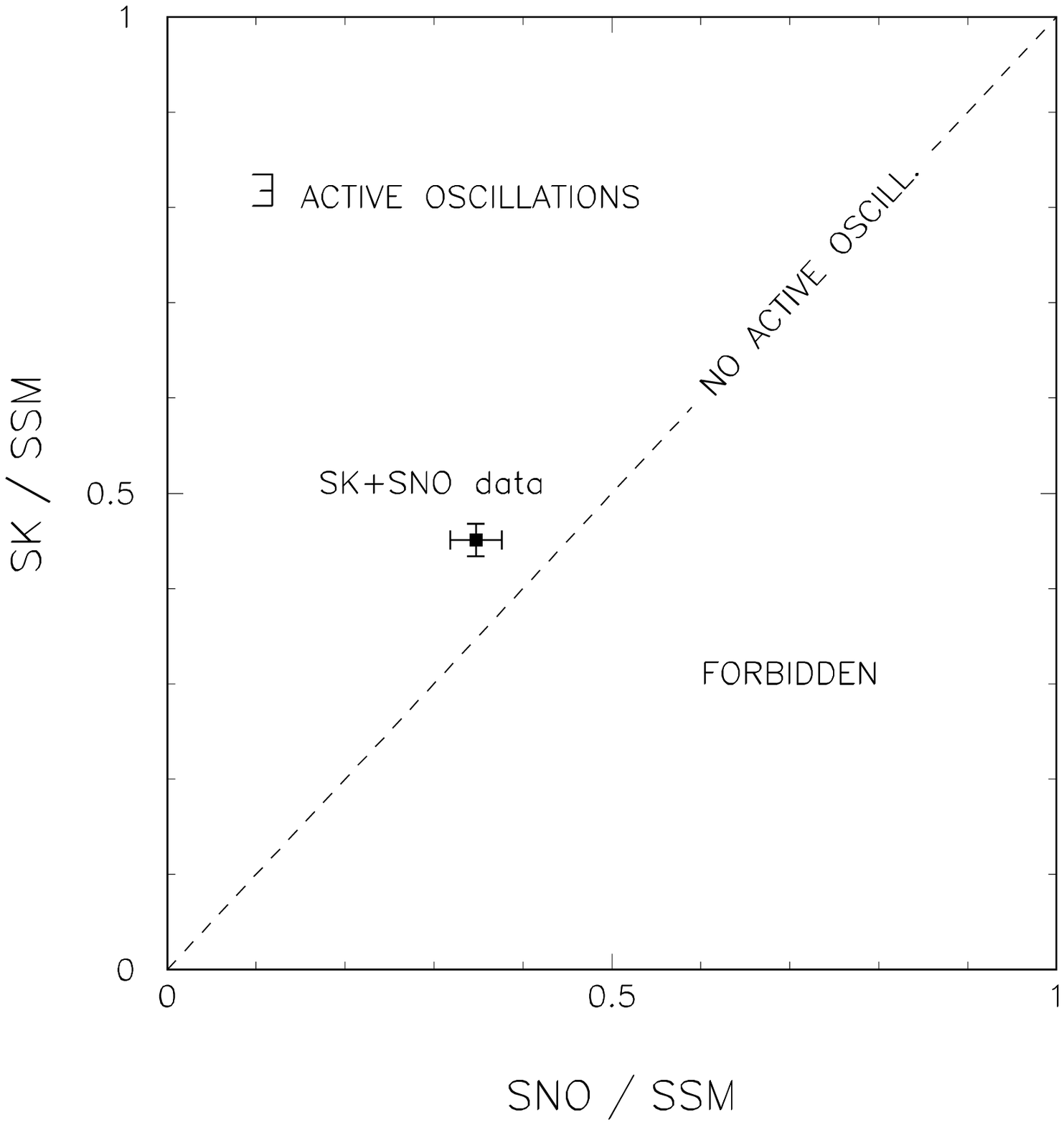}%
{Fig.~2. Model-independent consequences of the SK and SNO
results with equalized response functions: The data are well within the region
where active neutrino transitions $\nu_e\to\nu_{\mu,\tau}$ {\em must\/} occur,
and are $3.1\sigma$ distant from the diagonal line of  ``no active
oscillations''. This conclusion does not depend on either the standard solar
model or the possible presence of  additional transitions to sterile neutrinos.
See the text for details.}
\InsertFigure{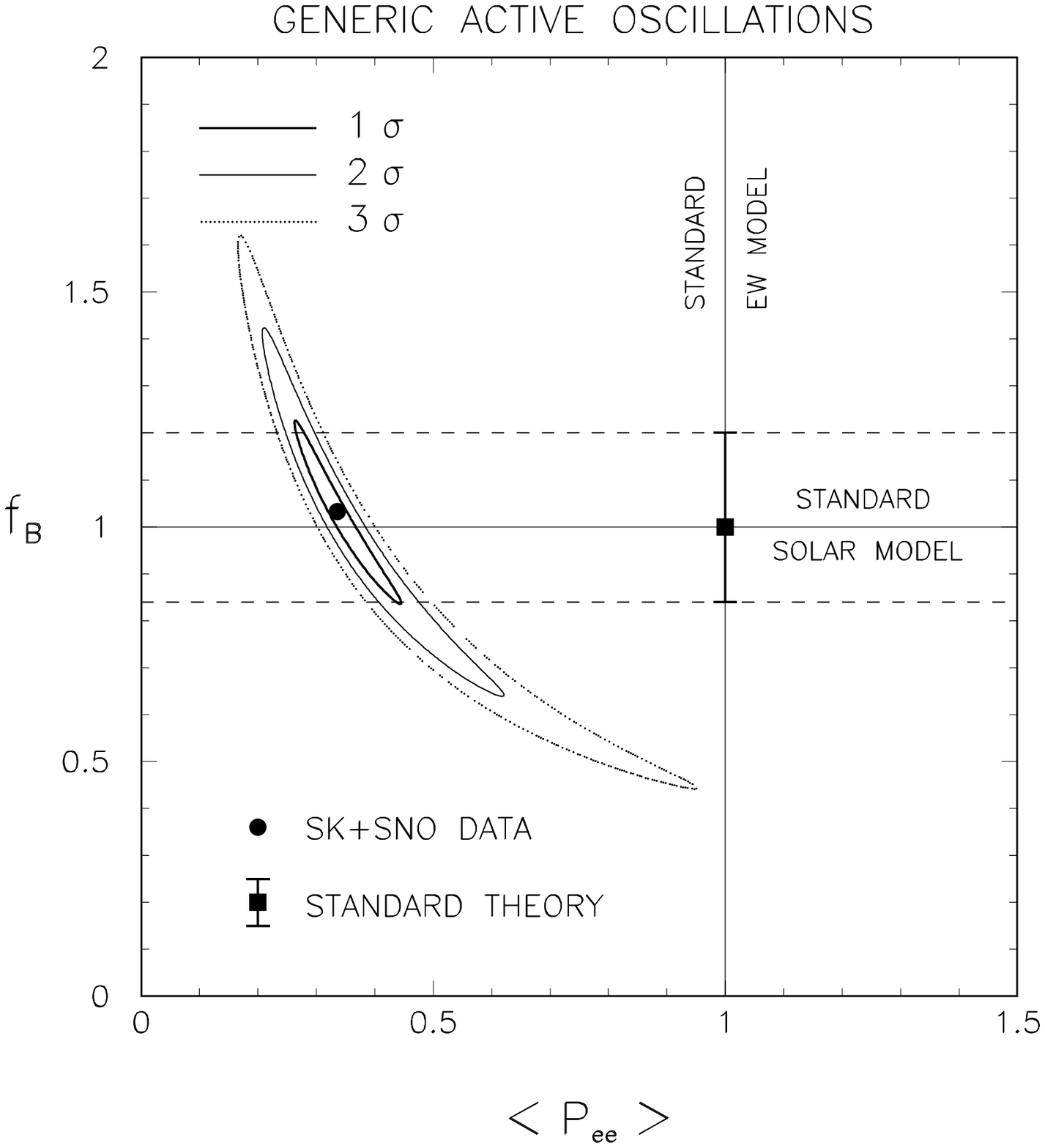}%
{Fig.~3. Model-independent analysis of SK and SNO, 
assuming no oscillations into sterile states, in the plane charted by $f_B$
(free factor multiplying the SSM $^8$B neutrino flux) and $\langle
P_{ee}\rangle$ ($\nu_e$ survival probability averaged over the SK-SNO response
function). The contours of the allowed region (obtained for $\Delta
\chi^2=1,\,4,$ and 9) give, after projections onto the axes ($N_{\rm DF}=1$),
the  separate $1\sigma$, $2\sigma$, and $3\sigma$ ranges for $f_B$ and $\langle
P_{ee}\rangle$. The $f_B$ range is in  good agreement with the SSM predictions
\cite{BP00} (shown as a $\pm 1\sigma$ horizontal band), while the $\langle
P_{ee}\rangle$ range is in $>3\sigma$ disagreement with the standard
electroweak model prediction of electron flavor conservation ($\langle
P_{ee}\rangle=1$).}
\InsertFigure{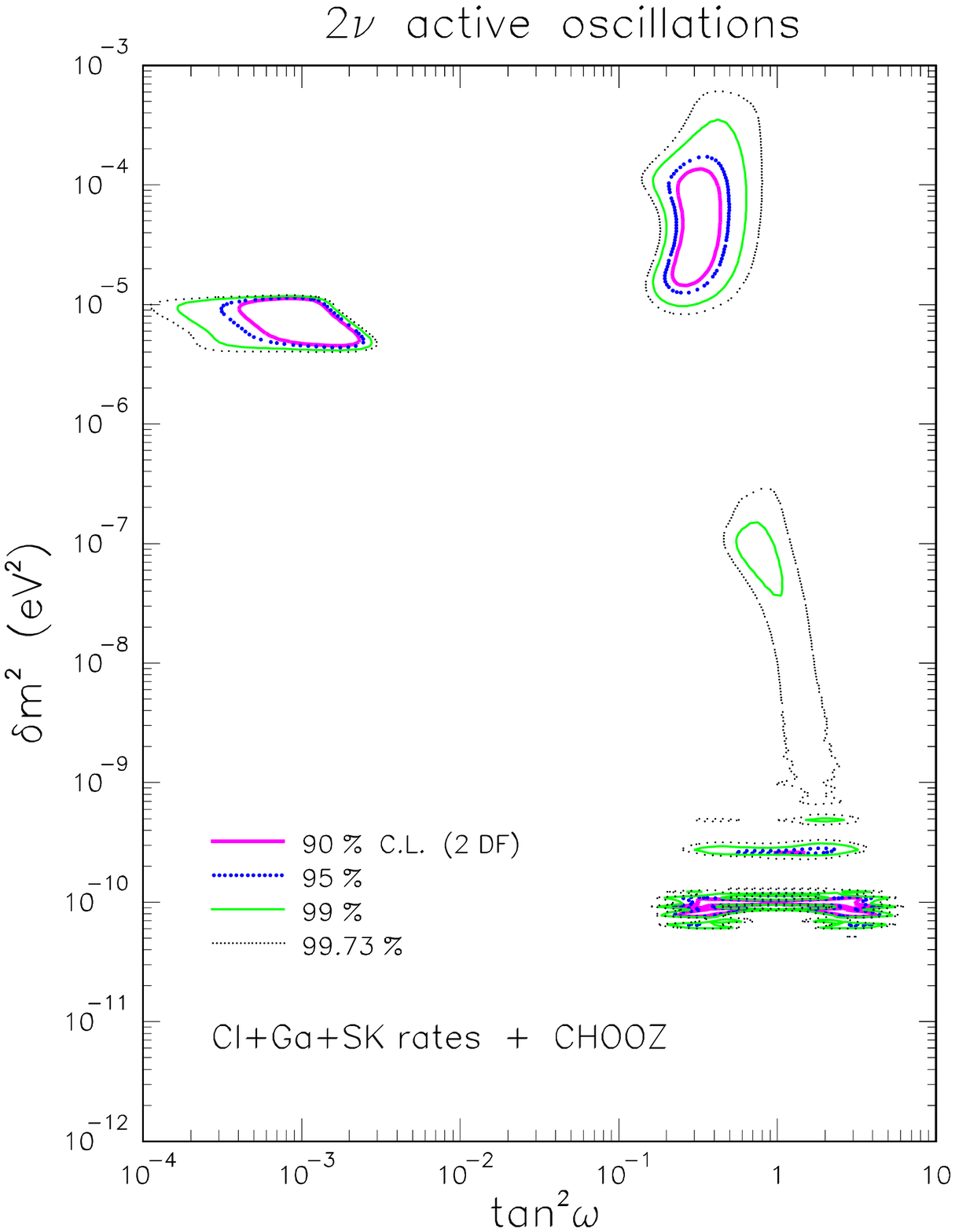}%
{Fig.~4. Pre-SNO $2\nu$ oscillation analysis of  total
neutrino event rates. CHOOZ data included. The regions shown in the figure are
allowed  90, 95, 99, and 99.73\% C.L.\ for the joint two-parameter estimation
test \cite{PDBS}, as obtained by drawing iso-$\chi^2$ contours at
$\Delta\chi^2=4.61,\,5.99,\,9.21,$ and 11.83 above the global minimum.}
\InsertFigure{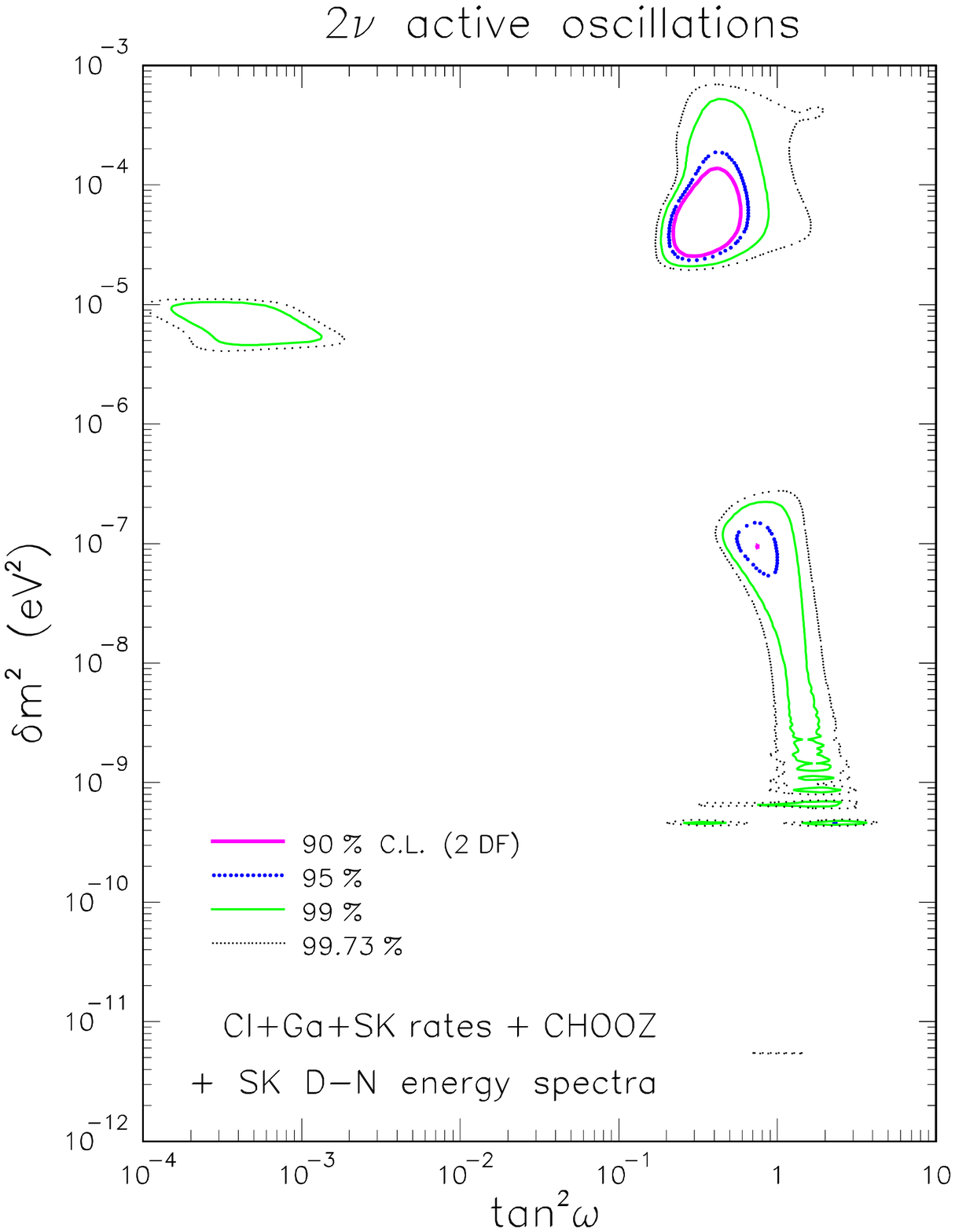}%
{Fig.~5. Pre-SNO  $2\nu$ oscillation analysis of total
neutrino event rates and of SK day-night energy spectra. CHOOZ data included.}
\InsertFigure{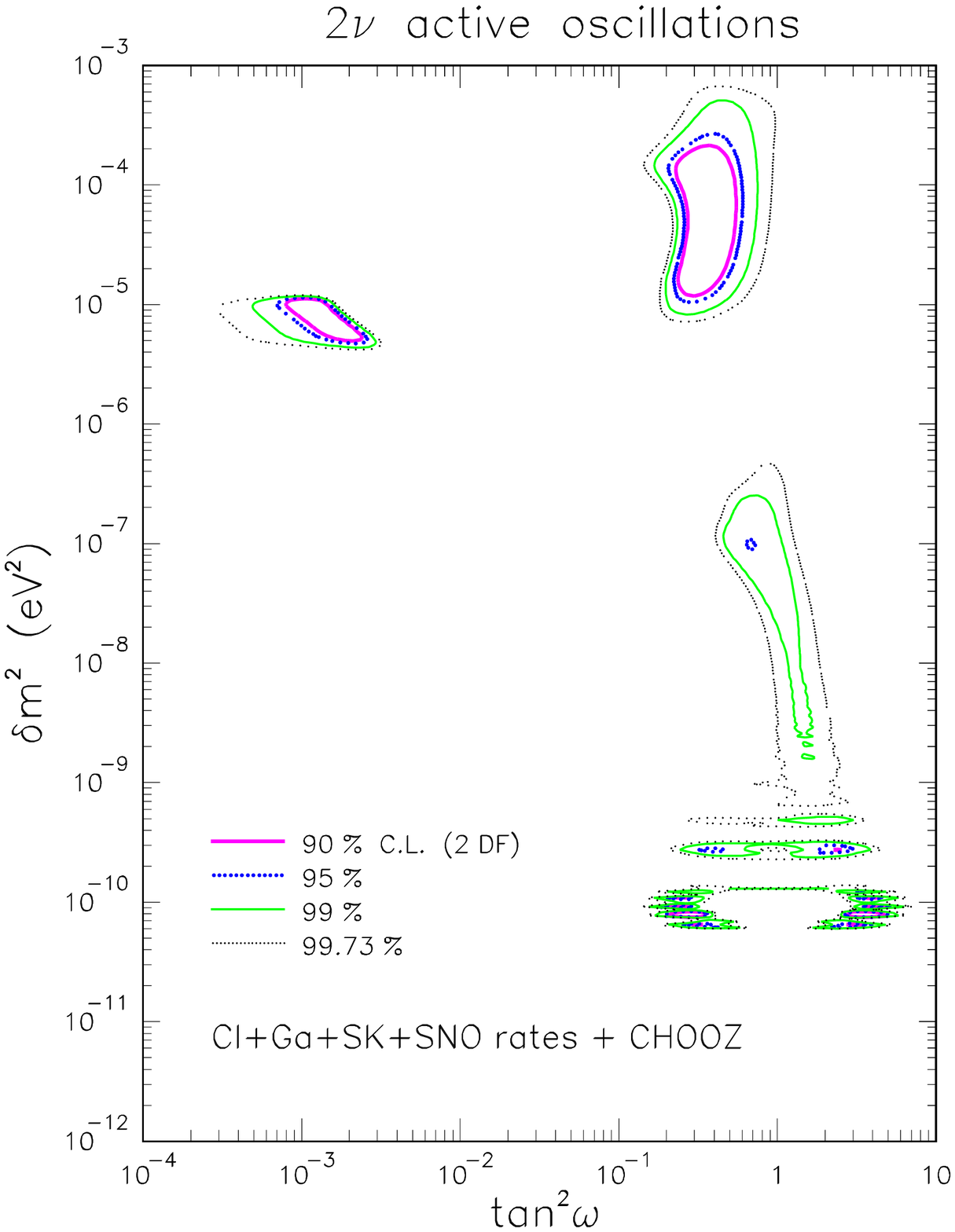}%
{Fig.~6. Post-SNO $2\nu$ oscillation analysis of  total
neutrino event rates. CHOOZ data included.}
\InsertFigure{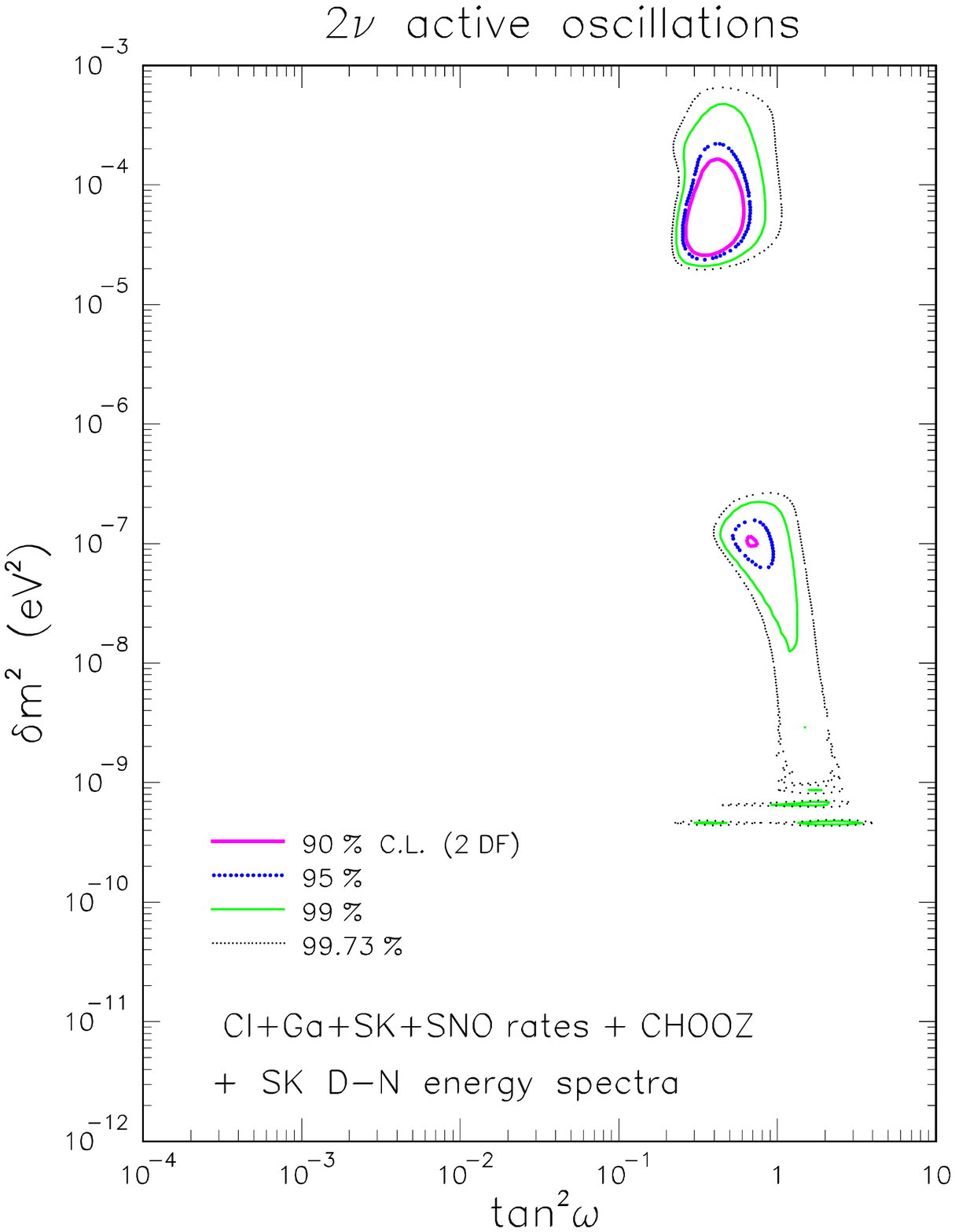}%
{Fig.~7. Post-SNO  $2\nu$ oscillation analysis of total
neutrino event rates and of SK day-night energy spectra. CHOOZ data included.
Solutions at small mixing are highly disfavored. See the text for details.}

\eject

\begin{thebibliography}{99}
\bibitem{FUTU}	We are currently revisiting several issues related to the 
		statistical treatment of (un)correlated errors in spectral
		measurements, that might lead to an improved estimate of
		bin-to-bin correlations. In the present work, however, we
		continue to use the official SK estimate of spectral
		uncertainties and their correlations \protect\cite{SK01,SKsp}.
\end{thebibliography}
\end{document}